\documentstyle[aps,preprint,prl]{revtex}

\tightenlines
\begin{document}

\title{Boson-conserving one-nucleon transfer operator in the
  interacting boson model} 

\author{J. Barea, C.E. Alonso, and J.M. Arias}

\address {Departamento de F\'{\i}sica At\'omica, Molecular y Nuclear,
  Universidad de Sevilla, \\ 
Apartado 1065, 41080 Sevilla, Spain}
\date{\today}
\maketitle

\begin{abstract}
The boson-conserving one-nucleon transfer operator 
in the interacting boson
model (IBA) is reanalyzed. Extra terms are
added to the usual form used
for that operator. These new terms change generalized 
seniority by one unit,
as the ones considered up to now. The results obtained 
using the new form for
the transfer operator are compared with those obtained 
with the traditional
form in a simple case involving the pseudo-spin 
Bose-Fermi symmetry $U^{B}(6) \otimes U^F(12)$ in its 
$U^{BF}(5) \otimes U^F(2)$ limit. 
Sizeable differences are
found. These results are of relevance in the study of 
transfer reactions to check nuclear supersymmetry and in 
the description of \( \beta  \)-decay within IBA. 
\end{abstract}

\vspace{2cm}

\noindent
{\bf PACS numbers: 21.60.Fw, 21.10.Jx, 23.40.Hc }
 
\newpage

The interacting boson model (IBA) has been extensively used in order to describe
spectroscopic 
properties of even-even and odd-even nuclei \cite{libroIBM,libroIBFM}. One
of the observables studied within IBA is the stripping and pick-up
single-nucleon transfer spectroscopic intensity 
between even-even and odd-even nuclei. The form of the one-nucleon
transfer operator in IBA is an open problem of current interest for several
reasons: i) some calculations indicate that the form commonly accepted
is insufficient 
for detailed calculations \cite{98Goll}, ii) it is one of the key points for
checking nuclear supersymmetry\cite{00Metz,01barea}, and iii) it is a
basic ingredient 
for the study of \( \beta  \)-decay since in IBA it is described as 
a composed process of a neutron (proton) pick-up followed by a proton (neutron)
stripping \cite{Fabio,89DI399,88ND2126}. Several forms 
for the one-nucleon transfer operator in IBA have been proposed in the
past based on different microscopic approaches. The original form,
proposed by Scholten and Dieperink \cite{81Sc343,85Sc189}, 
was obtained  in the generalized seniority \( (\tilde{v}) \) 
scheme \cite{71Ta1} by using the Otsuka-Arima-Iachello 
(OAI) mapping method \cite{79Ot93} and the number approximation
\cite{Otsuka}. 
This is still the operator implemented in the standard IBA
codes and the most commonly used. Alternatively, other forms based on
the nuclear field theory techniques \cite{89So317}, the use of RPA for
quadrupole phonons and mapping the fermion space onto the
fermion-quadrupole space \cite{84PB143}, 
and the generalized Holstein-Primakoff
mapping method \cite{92Al391,94DP291} have been proposed but there 
have been few actual calculations using these operators. 

In this paper we reanalyze the standard 
IBA boson-conserving one-nucleon transfer operator proposed originally
by Scholten  
and Dieperink and improve that operator by including extra terms which
change 
seniority by one unit and are the next order in $d-boson$ creation and
annihilation 
operator. We show that this new type of terms give rise to sizeable
changes in 
the one-nucleon transfer spectroscopic intensities and consequently
conclude that they must be included in the standard IBA operator.

The starting point is the IBA boson-conserving part of the 
one-nucleon transfer operator 
derived semi-microscopically by Scholten and Dieperink
\cite{81Sc343,85Sc189} using the generalized seniority 
formalism \cite{71Ta1}. 
Consequently, it is only strictly valid for spherical nuclei where
seniority is approximately a good quantum number. However, this operator 
has also been used with reasonable success 
for transfer reactions involving deformed 
nuclei \cite{98Goll,00Metz}. 
Scholten and Dieperink constructed the Interacting Boson-Fermion
Approximation (IBFA) image of the shell model single-nucleon creation 
operator, taking into account 
terms which change generalized seniority by one unit and using fermion
states with \( \tilde{v}\leq 2 \). Under these approximations the transfer
operator which does not change the number of bosons is

\begin{equation}
\label{opviejo}
{\mathcal{T}}_{jm}^{\dagger }=u_{j}\, a_{jm}^{\dagger }-\sum _{j^\prime }\frac{v_{j}}{\sqrt{N}}\beta _{j{^{\prime }}j}\sqrt{\frac{10}{2j+1}}[(s^{\dagger }\times \tilde{d})^{(2)}\times a^{\dagger }_{j^\prime }]^{(j)}_{m}\, ,
\end{equation}
where \( a_{jm}^{\dagger } \) is the odd-nucleon creation operator in the
IBFA boson-fermion space, \( v_{j} \) is the occupation probability of the
single particle orbit \( j \) and \( u_{j}=\sqrt{1-v_{j}^{2}} \). \( N \)
is the number of bosons. The summation on \( j' \) runs over the
valence single particle angular momenta, and the normalized quantities 
\( \beta_{j'j} \) are \cite{81Sc343,85Sc189}

\begin{equation}
\beta _{j'j}=Q_{j'j}(u_{j'}v_{j}+v_{j'}u_{j})/N_{\beta }\, ,
\label{betas}
\end{equation}
where
\begin{equation}
Q_{j'j}=\langle  j'||Y^{(2)}|| j\rangle
\label{Qus}
\end{equation}
and \( N_{\beta } \) is a normalization constant obtained from the condition

\begin{equation}
\sum _{jj'}(\beta _{jj'})^{2}=1
\end{equation}
where \( j \) and \( j' \) are the valence single particle 
angular momenta. The operators
\( \tilde{d}_{m} \) , which behave appropriately under rotations, are
\( \tilde{d}_{m}=(-1)^{m} d_{-m} \) .

As mentioned before, the expression (\ref{opviejo}) was obtained under the
restrictions of changing 
generalized seniority by one unit and using fermion states with 
\( \tilde{v} \leq 2 \). The term 
\( [(d^{\dag }\times \tilde{d})^{(\lambda )}\times a_{j'}^{\dag }]_{m}^{(j)} \)
changes generalized seniority in one unit as 
\( [(s^{\dag }\times \tilde{d})^{(2)}\times a_{j'}^{\dag }]^{(j)}_{m}\), 
and does not appear in the traditional expression of the 
one-particle transfer operator. The reason is that the derivation of
the corresponding coefficient 
implies the use of fermion states with \( \tilde{v} > 2\). 
However, this kind 
of terms are obtained naturally when using nuclear field theory techniques
\cite{89So317}, or the generalized Holstein-Primakoff mapping method
\cite{92Al391}. 
For this reason it is natural to wonder which is the influence of this
kind of terms.
Taking into account these terms, the new transfer operator reads
\footnote{We prefer to write the operator in this way but with the
  appropriate angular momenta recoupling it can be written as
$$
{\mathcal{T}}^{\dagger \prime }_{j m} = {\mathcal{T}}^{\dagger
  }_{j m} - \sum _{j^\prime ,J,\lambda }\, (-1)^{\lambda }\, \widehat{J}\widehat{\lambda }\left\{ \begin{array}{ccc}
2 & 2 & \lambda \\
j & j^\prime & J
\end{array}\right\} \phi _{j j^\prime }^{J}[(d^{\dagger }\times
\tilde{d})^{(\lambda )}\times a^{\dagger }_{j^\prime }]^{(j)}_{m}. 
$$}

\begin{equation}
{\mathcal{T}}^{\dagger \prime }_{j m}={\mathcal{T}}^{\dagger }_{j
  m}+\sum _{j^\prime,J}\, \phi ^{J}_{j j^\prime }[(a_{j^\prime}^{\dagger }\times d^{\dagger })^{(J)}\times \tilde{d}]^{(j)}_{m}. 
\label{opnuevo}
\end{equation}
This is the operator to be used in IBFA if no distinction between
protons and neutrons is made. When working with versions
of the IBA--IBFA in 
which the proton-neutron degrees of freedom are treated explicitly,
all the 
operators in the preceding equation will have an index 
\( \pi  \) (\( \nu ) \) 
if the transferred nucleon is a proton (neutron). 
In this work the coefficients \( \phi ^{J}_{j j^\prime } \)
will be obtained by requiring that the matrix elements of the fermion
operator  $C^\dagger_j$ and the IBFA operator 
${\mathcal{T}}^{\dagger \prime  }_j$ between 
states of $\tilde{v}=2$ and $\tilde{v}=3$ in the corresponding spaces
are equal. 
For that purpose, first one has to construct the $\tilde{v}=3$ shell
model states which, in general, will not be orthonormal. From this set
of non-orthonormal fermion states a mapping procedure onto the
$\tilde{v}=3$ orthonormal boson states is needed. In this paper we
use the so called {\it democratic} mapping
\cite{demo}. This mapping is based on the diagonalization of the shell
model overlap matrix. In our case, the method starts from
a set of $n$ fermion states which are linearly independent but not necessarily
orthonormal $|F,(i);JM\rangle $ 
($i=1,\dots,n$) and a
set of $n$ orthonormal boson states $|B,(i);JM\rangle $
($i=1,\dots,n$). The {\it democratic} mapping follows the following steps:

\begin{enumerate}
\item First, one has to construct the initial fermion states 
 $|F,(i);JM\rangle $  ($i=1,\dots,n$) 
 and the orthonormal boson states $|B,(i);JM\rangle $
($i=1,\dots,n$). 
 
\item  Second, one has to construct the matrix $\Theta^J$ with the
  overlaps of the fermionic states for
  each total angular momentum $J$,
\begin{equation} 
\Theta^J_{i k}=\langle F,(i);JM|F,(k);JM\rangle ~;~ i,k=1,\dots,n
.
\end{equation}

\item The diagonalization of the matrix $\Theta^J$ provides its
  eigenvectors and eigenvalues. Let us call $C^J$ the $n\times n$ square
  matrix containing the
  eigenvectors (in columns) fixed by the condition $\sum_k (C^J_{i
  k})^2=1$ and ${\mathcal{\lambda}}^J$ the diagonal matrix
  containing the eigenvalues, $\lambda_i^J$. The following
  relations hold in matrix
  notation 
\begin{equation}
\Theta^J \cdot C^J = C^J \cdot {\mathcal{\lambda}}^J
\end{equation}
and 
\begin{equation}
\left(C^J\right)^T = \left(C^J\right)^{-1} . 
\label{orto}
\end{equation}
The orthonormal eigenvectors are,
\begin{equation}
|F,k;JM\rangle_{\perp }= {1 \over \sqrt{\lambda_k^J}} \sum_{i=1}^n C^J_{i,k}
|F,(i);JM\rangle ~; ~ k=1,\dots,n .
\label{Fok}
\end{equation}

\item The boson states are transformed to a new orthonormal set
  according to the same transformation as the fermion states
\begin{equation}
|B,k;JM\rangle_{\perp }= \sum_{i=1}^n C^J_{i,k} |B,(i);JM\rangle ~ ; ~
k=1,\dots,n .
\label{Bok}
\end{equation}
It is easy to see that these new boson states are orthonormal using
Eq. (\ref{orto}).

\item Finally, the mapping is done between the new fermion,
  Eq. (\ref{Fok}), and boson, Eq. (\ref{Bok}), states
\begin{equation}
|F,k;JM\rangle_{\perp } \Longleftrightarrow |B,k;JM\rangle_{\perp} ~ ;  ~ k=1,\dots,n.
\end{equation}

\end{enumerate}

In the following, we proceed with these steps and derive
semi-microscopically, in the same spirit as Scholten and Dieperink,
the coefficients \( \phi ^{J}_{j j^\prime } \).

First, it is necessary to construct properly the
\( \tilde{v}=3 \) states in the shell model space. 
In order to get this set, we start with the following
unnormalised shell model states 
\begin{equation}
\label{v3+v1}
|S^{\, N-1}(Dj);\, JM\rangle=[C_{j}^{\dagger }\times D^{\dagger }]_{M}^{J}\, (S^{\dagger })^{N-1}|0\rangle,
\end{equation}
which, in general, contain \( \tilde{v}=1 \) and \( \tilde{v}=3 \)
components. \( C_{j}^{\dagger } \) represents the creation operator of
one nucleon in the 
single particle orbit \( j \)  and the fermion pair creation operators 
\( S^{\dagger }\, \textrm{and}\, D^{\dagger } \) are \cite{81Sc343},
\begin{equation}
S^{\dagger } = \sum_j \alpha_j \sqrt{{\Omega_j \over 2}}
(C_{j}^{\dagger } C_{j}^{\dagger })^{(0)}
\end{equation}
and
\begin{equation}
D^{\dagger } = \sum_{j j^\prime} \beta^\prime_{j j^\prime} \sqrt{{1 \over 2}}
(C_{j}^{\dagger } C_{j^\prime}^{\dagger })^{(2)}
\end{equation}
where the structure coefficients $\alpha_j$ and 
$\beta^\prime_{j j^\prime}$ are,
\begin{equation}
\alpha_j= \sqrt{{\Omega_e \over N}} v_j,
~~~
\beta^\prime_{j j^\prime}= { \beta_{j j^\prime} \over u_j u_{j^\prime}}~,
\end{equation} 
where $\beta_{j j^\prime}$ are defined in 
Eq. (\ref{betas}),
\( \Omega _{j} = (2j+1)/2 \) and 
\( \Omega _{e} = \sum_j \alpha_j^2 (2j+1)/2 \) is the effective
shell degeneracy\cite{81Sc343}. 

Unnormalized \( \tilde{v}=3 \) states are constructed
subtracting the \( \tilde{v}=1 \) components in (\ref{v3+v1}) in the
following way 
\begin{equation}
|\tilde{v}=3;\, S^{\, N-1} (Dj);J\, M\rangle=\frac{|S^{\,
    N-1} (Dj);\, JM\rangle}{u_j \sqrt{\frac{(N-1)!\Gamma (\Omega
      _{e}+1)}{\Gamma (\Omega _{e}+1-(N-1))}}}-\chi _{j}^{J}
    |\tilde v=1, N; JM \rangle,
\label{v3norto}
\end{equation}
where the denominator in the first term on the right-hand-side is
introduced for later convenience in order to simplify the expressions
of some quantities given below. $|\tilde v=1, N; JM \rangle $ are
properly normalized seniority $\tilde v=1$ states. 
\( \chi ^{J}_{j} \) is obtained
requiring that the overlap between the states (\ref{v3norto}) and the
states of seniority one is zero. The expression obtained is 
\begin{equation}
\chi _{j}^{J}=-\sqrt{\frac{10}{2J+1}}\, \frac{v_{j}}{u_{j}}\beta _{Jj}\, ,
\label{17}
\end{equation}
if \( J \) is equal to one of the allowed angular momenta, and zero
otherwise. In the above derivation the number approximation 
\cite{Otsuka} 
$\langle 0 | S^N (S^\dagger)^N |0 \rangle = \frac{N!\Gamma (\Omega
      _{e}+1)}{\Gamma (\Omega _{e}+1-N)}$ has been used.

Once we have the non-orthonormal fermion states,
Eqs. (\ref{v3norto}) and (\ref{17}),
the matrix of
the overlaps, \( \Theta ^{J} \), can be constructed, obtaining

\begin{eqnarray}
\nonumber
\Theta _{mr}^{J} & = & \langle\tilde{v}=3;S^{\, N-1}(Dj_{m});\, JM\, |\tilde{v}=3;S^{\, N-1}(Dj_{r});\, JM\rangle  \\
 & = & \delta _{mr}+10\left( \sum ^{n}_{i=1}\delta _{j_{i}\, ,J}\right) \sum ^{n}_{k=1}\beta _{j_{k},j_{r}}\beta _{j_{k},j_{m}}\left\{ \begin{array}{ccc}
j_{r} & 2 & j_{k}\\
j_{m} & 2 & J
\label{overlaps}
\end{array}\right\} .
\end{eqnarray}
Note that this expression tells us that the states (\ref{v3norto}) 
are already orthonormal when \( J \) does not coincide with
one of the valence single particle angular momenta. 
For the general case in which there are several single particle
angular momenta $j$ that contribute  to a given $J$ one has to follow
the procedure sketched above. First, the matrix $\Theta^J$ is
diagonalized obtaining its eigenvectors, $C^J$, and eigenvalues,
$\lambda^J$. Then, orthonormalized fermion and boson seniority three states are
constructed as given in Eqs. (\ref{Fok}) and (\ref{Bok})
\begin{equation}
|F,\tilde{v}=3,k;JM\rangle_{\perp}= {1 \over \sqrt{\lambda_k^J}} \sum_{i=1}^n C^J_{i,k}
|\tilde{v}=3;S^{\, N-1}(Dj_{i});\, JM\rangle ~ ; ~ k=1,\dots,n .
\end{equation}
and
\begin{equation}
|B,\tilde{v}=3,k;JM\rangle_{\perp}= \sum_{i=1}^n C^J_{i,k} | s^{N-1}(d j_{i});\,
JM \rangle ~ ; ~ k=1,\dots,n .
\end{equation}
We are now in a position to map
the \( \tilde{v}=3 \) shell model states into the \( \tilde{v}=3 \)
boson-fermion states of the IBFA 
\begin{equation}
|F,\tilde{v}=3,k;JM\rangle_{\perp} \Longleftrightarrow
|B,\tilde{v}=3,k;JM\rangle_{\perp}, ~ k=1,\dots,n .
\end{equation}
Since now the \( \tilde{v}=3 \) states are known both in the fermion
space and in the boson-fermion space, 
the coefficients \( \phi ^{J}_{j_{k}j_{k^\prime }} \)
will be obtained by requiring that the matrix elements of the fermion operator 
$C^\dagger_{j_k}$ and the IBFA operator ${\mathcal{T}}_{j_k}^{\dagger
  \prime }$ between  
states of $\tilde{v}=2$ and $\tilde{v}=3$ in the corresponding spaces are equal,  
\begin{equation}
_{\perp}\langle F,\tilde{v}=3, \ell;J
||C^{\dagger }_{j_{k}}||S^{N-1}D;2\rangle=_{\perp}\langle B,\tilde{v}=3,\ell;J
||{\mathcal{T}}^{\dagger \prime }_{j_{k}}||s^{N-1} d;2\rangle.
\end{equation}

With this requirement the following expression for the coefficients 
\( \phi ^{J}_{j_{k}j_{k^\prime }} \) is obtained

\begin{equation}
\phi _{j_{k}j_{k^\prime}}^{J}=(-1)^{j_{k}-j_{k^\prime \,
    }}u_{j_{k}}\widehat{\frac{J}{\widehat{j_{k}}}} \left[\delta
    _{j_{k}j_{k^\prime }}-\left(C^J \cdot \sqrt{{\mathcal {\lambda}}^J} \cdot
    (C^J)^T\right)_{k k^\prime}\right],
\label{coef}
\end{equation}
where \( \widehat{j}=\sqrt{2j+1} \) and \( \sqrt{{\mathcal {\lambda}}^J} \) is
a diagonal matrix whose elements are the square root of of the
corresponding ones in $\lambda^J$ (the square root of the eigenvalues
of $\Theta^J$).
The transfer operator given in Eq. (\ref{opnuevo}) with the coefficients in
Eq. (\ref{coef}) is the form proposed in 
this work for a particle-like transfer.
For the case of a hole-like transferred nucleon 
the \( u's \) and \( v's \) have to be interchanged 
(in addition it should be noted that the reduced matrix elements given in 
Eq. (\ref{Qus}) change sign if the transferred nucleon is hole-like\cite{BM}). 
This new form for the boson-conserving one nucleon transfer operator
has been implemented in the IBA codes \cite{Roelof} which are
available under request.  

We have performed calculations both in IBA-1 and IBA-2
with these new terms
in order to show their relative importance and as a result sizeable
differences are found. Below we present a simple example.

The observable we calculate is the spectroscopic intensity defined as,

\begin{equation}
\label{oe-ee}
I_{j}(J_{i}\rightarrow J_{f})=|\langle J_{f}||{\mathcal{T}}^{\dagger \prime }_{j}||J_{i}\rangle|^{2}.
\end{equation}
$J_i$ is a state in the even-even nucleus and $J_f$ a state in the
odd-even. 
%The operator corresponding to a transfer from the odd-even
%to the even-even nucleus is $\tilde {\mathcal{T}}_j$. Thus
It is easy to show that 
$I_{j}(J_{i}\rightarrow J_{f}) = I_{j}(J_{f}\rightarrow J_{i})$ and 
this fact allows us not to specify which is the initial
nucleus in the table presented below.

As a simple example we present here 
one-nucleon particle-like transfer intensities
between two nuclei which follow particular dynamical symmetries of the
IBA-1 and IBFA-1. 
The even-even nucleus has just one boson, 
is the boson core for the odd-even nucleus. 
%and follows the \( U^{B}(5) \) dynamical symmetry. 
The odd-even nucleus 
has one boson plus one particle and is described within the pseudo-spin
Bose-Fermi dynamical symmetry \( U^{BF}(5) \otimes U^F(2) \)
\cite{84PAS183} obtained from the product \( U^{B}(6)\otimes U^{F}(12) \),
where the superscripts \( B \) and \( F \) stand for boson and fermion.
In the odd-even nucleus, the odd particle has been allowed to occupy the 
\( j=1/2,3/2 \) , and \( 5/2 \) single particle states. 
Under these conditions, once the number of bosons is fixed, the wave
functions are 
fixed by the symmetry, independently of the parameters used in the
Hamiltonian. 
The occupation probabilities that appear in the transfer operator are
not provided by the symmetry. In the example presented here the values
used are,
 \( v_{1/2}^{2}=0.5,\, \, v_{3/2}^{2}=0.25 \) and \( v_{5/2}^{2}=0 \) which 
fulfil \( \sum _{j_{k}}(2j_{k}+1)v_{j_{k}}^{2}=2N \).

The states in the  even-even core nucleus are identified by the
labels of the irreducible representations of the group chain 
\( U^{B}(6)\supset U^{B}(5)\supset O^{B}(5)\supset O^{B}(3) \):
\( |[N] \langle n_{d}\rangle (v) L\rangle \). 
In our case, with just one boson, there are
only two states, the ground state 
\( |0_B\rangle= \)\( |[1] \langle 0\rangle (0) 0\rangle \) with
one \( s-boson \) and the excited state 
\( |2_B\rangle= \)\( |[1] \langle 1\rangle (1) 2\rangle \) with
one \( d-boson \). The states in the odd-even nucleus are labelled by
the labels of the irreducible representations of the group chain 
\( U^{BF}(6)\supset U^{BF}(5)\supset O^{BF}(5)\supset O^{BF}(3)\supset Spin^{BF}(3) \):
\( |[N_{1},N_{2}] \langle n_{1},n_{2}\rangle (v_{1},v_{2})L;J\rangle \).

Since the ground state
of the even-even nucleus has no \( d-bosons \) the only term of the
transfer 
operator that contributes to the transfers from this state to any
state in the 
final odd-even nucleus is the \( a_{j_{k}}^{\dagger } \) term. We have
checked that 
our results in this case coincide with those obtained analytically in
Ref. \cite{84PAS183}.
The calculated spectroscopic intensities
between the state  \( |2_B\rangle \)
in the even-even nucleus and all possible states in its odd-even
counterpart are presented in Table \ref{tabla10}. 
In the first column the quantum labels corresponding to
the state 
in the odd-even nucleus are specified. Two calculations are shown, the
first one under 
the label \( gs \) for the traditional operator given in
Eq. (\ref{opviejo}),  and the second
one under the label \( gs+ \) for the full operator proposed in this work,
Eqs. (\ref{opnuevo}) and (\ref{coef}). 
Important differences between both calculations can be seen for some
transfers. Let us discuss briefly the transfers involving the state \(
|2_B\rangle \) 
in the even-even nucleus and the states 
\( |[2,0] \langle 0,0\rangle (0,0) 0;1/2\rangle \) and 
\( |[1,1] \langle 1,1\rangle (1,1) 1;1/2\rangle\)
 in the odd-even
nucleus.
The state  \( |[2,0] \langle 0,0\rangle (0,0) 0;1/2\rangle \) has just
one component \( |(0_B \times a^\dagger_{1/2})^{(1/2)}\rangle \)
 (see table XIII in Ref. \cite{84PAS183}), thus the connection
with the state \( |2_B \rangle \) in the even-even nucleus can only be
done through the terms 
\( (s^\dagger \tilde d a^\dagger_{1/2})^{(3/2,5/2)} \) in the
transfer operator. The new terms proposed in this work
give no contribution to these
transfers as it is seen in Table \ref{tabla10}. The value zero for the
$j=5/2$ 
transfer is due to the value $v_{5/2}^2=0$ selected. On the other
hand, the state \( |[1,1] \langle 1,1\rangle (1,1) 1;1/2\rangle\) has
two components (see table XIII in Ref. \cite{84PAS183}): 
$\sqrt{3/5}  |(2_B \times a^\dagger_{3/2})^{(1/2)}\rangle +
\sqrt{2/5}  |(2_B \times a^\dagger_{5/2})^{(1/2)}\rangle $.
The connection with the state \( |2_B \rangle \) in the even-even
nucleus can be done only by the $a^\dagger_{j=3/2,5/2}$ and the new terms.
The values under the label $gs$ are those coming
just from $a^\dagger_j$. The values under the label $gs+$ include the
contributions from $a^\dagger_j$ and those coming from 
the proposed terms $ (d^\dagger \tilde d a^\dagger)$. It can be seen
that these contributions are important, constructive in one case and
destructive in the other case. 

It can be in Table \ref{tabla10} 
observed that some forbidden transfers with the traditional operator
given in Eq. (\ref{opviejo}) become allowed with the new one, Eq. (\ref{opnuevo}).
We have performed calculations with larger number of bosons 
with IBA-1 and IBFA-1 and with the two types of bosons
(neutron and proton) (IBA-2,IBFA-2) and in every case sizeable
differences are observed.

In conclusion, we have calculated extra \( \Delta \widetilde{v}=1 \) terms
in the bosonic expansion of the transfer operator which connects
states of IBA
with states of IBFA. Through the comparisons between the transfer
intensities 
calculated using the transfer operator with and without these terms in 
the \( U^{BF}(5) \) limit of the pseudo-spin symmetry 
\( U^{BF}(6) \otimes U^F(2) \)\cite{84PAS183}, we have shown that
the added terms are not negligible when compared to the ones taken into account up
to now and expect improvements in the description of the one nucleon
transfer intensities 
using this modified operator. Calculations in the Pt region are in progress
and will we published somewhere else \cite{unpub}.

The influence of this modified operator for the calculation of
beta-decay within the interacting boson model remains to be studied. 
The matrix elements needed in this case are of the same kind
as the ones needed in order to calculate
spectroscopic intensities. However, in beta-decay all transfers
connecting an initial state in an odd-even nucleus 
(not necessarily the ground state) to all
possible states in the even-even intermediate nucleus and then all
transfers from these intermediate states to the final state (not
necessarily the ground state too) 
in the even-odd nucleus have to be calculated. Thus, many transfers
contribute and it is important to study the changes induced by the new transfer
operator in its description.

Work supported in part by Spanish CICYT under contract PB98-1111.

\begin{table}
\caption
{Calculated spectroscopic intensities for one-nucleon
transfer between an even-even nucleus with $ N=1 $
and an odd-even \protect\protect\( U^{BF}(5)\protect \protect \)
nucleus with $ N=1 $ plus one fermion. 
Results for transferred $j=1/2$, $j=3/2$, and 
$j=5/2$ are shown. Calculations made with 
(\ref{opviejo}) and (\ref{opnuevo}) are displayed under the labels
\( gs \) and \( gs+ \), respectively. Only transfers involving the state 
\( |2_B \rangle=|{[}1{]}\langle 1\rangle (1)2\rangle  \) 
in the even-even core 
nucleus are included, since those involving the state 
\( |0_B \rangle=|{[}1{]}\langle 0\rangle (0)0\rangle  \) 
in the even-even core 
have no contribution from the added terms to the transfer operator. The quantum 
numbers in the first column specify the state in the odd-even nucleus.}
\vspace{0.5cm} 
\begin{center} 
\begin{tabular}{c|cc|cc|cc}
 $[N_1,N_2] \langle n_1,n_2 \rangle (v_1,v_2)  L;  J$ &\multicolumn{2}{c|}{j=1/2} &\multicolumn{2}{c|}{j=3/2} 
&\multicolumn{2}{c}{j=5/2} \\ 
 odd-even state & $gs$ & $gs+$ & $gs$ & $gs+$ & $gs$ & $gs+$ \\
\hline
$[1,1] \langle 1, 1\rangle (1,1)  1;  1/2$ &  &  & 0.900 & 0.477 & 0.800 & 1.542 \\
$[2,0] \langle 0, 0\rangle (0,0)  0;  1/2$ &  &  & 0.539 & 0.539 & 0.000 & 0.000 \\
$[2,0] \langle 2, 0\rangle (0,0)  0;  1/2$ &  &  & 0.600 & 0.980 & 1.200 & 0.844 \\
$[1,1] \langle 1, 1\rangle (1,1)  1;  3/2$ & 0.000 & 0.007 & 0.900 & 1.638 & 2.800 & 2.181 \\
$[1,1] \langle 1, 0\rangle (1,0)  2;  3/2$ & 3.007 & 2.526 & 0.217 & 0.067 & 0.000 & 0.003 \\
$[2,0] \langle 1, 0\rangle (1,0)  2;  3/2$ & 0.071 & 0.015 & 0.217 & 0.451 & 0.000 & 0.003 \\
$[2,0] \langle 2, 0\rangle (2,0)  2;  3/2$ & 0.000 & 0.053 & 2.100 & 1.659 & 1.200 & 2.164 \\
$[1,1] \langle 1, 0\rangle (1,0)  2;  5/2$ & 3.546 & 5.168 & 0.031 & 0.220 & 0.000 & 0.027 \\
$[1,1] \langle 1, 1\rangle (1,1)  3;  5/2$ & 0.000 & 0.142 & 3.600 & 3.217 & 1.200 & 2.441 \\
$[2,0] \langle 1, 0\rangle (1,0)  2;  5/2$ & 0.321 & 0.916 & 0.031 & 0.014 & 0.000 & 0.027 \\
$[2,0] \langle 2, 0\rangle (2,0)  2;  5/2$ & 0.000 & 0.000 & 0.900 & 1.662 & 4.800 & 4.921 \\
$[1,1] \langle 1, 1\rangle (1,1)  3;  7/2$ &  &  & 0.600 & 0.600 & 7.200 & 7.200 \\
$[2,0] \langle 2, 0\rangle (2,0)  4;  7/2$ &  &  & 5.400 & 5.400 & 0.800 & 0.800 \\
$[2,0] \langle 2, 0\rangle (2,0)  4;  9/2$ &  &  &  &  & 10.000 & 10.000 \\
\end{tabular}
\end{center}
\label{tabla10}
\end{table}

\end{document}